\documentstyle[prl,aps,twocolumn,psfig,epsfig,subfigure]{revtex} 
\tighten 
\newcommand{\bcols}{\ifpreprintsty\else\begin{multicols}{2}\fi}
\newcommand{\ecols}{\ifpreprintsty\else\end{multicols}\fi}

\begin{document}
\bibliographystyle{prsty} 
\draft

\title{An `all-carbon' molecular switch}
\author{R. Gutierrez$^a$\thanks{email: gutie@theory.phy.tu-dresden.de},
G. Fagas$^b$, G. Cuniberti$^b$, F. Grossmann$^a$, R. Schmidt$^a$, and K. Richter$^c$}
\address{
$^a$Institute for Theoretical Physics , Technical University of Dresden, D-01062 , Germany\\
$^b$ Max Planck Institute for the Physics of Complex Systems, N{\"o}thnitzer
Str. 38,
D-01187 Dresden,Germany\\
$^c$ Institute for Theoretical Physics, University of Regensburg,
D-93040 Regensburg,Germany}

\maketitle

\begin{abstract}
We have performed parameter-free calculations of electron transport across  a
carbon molecular junction consisting of a C$_{60}$ molecule
sandwiched between two semi-infinite metallic carbon nano\-tubes. 
It is shown that the Landauer conductance of this carbon hybrid system
can be tuned within orders of magnitude not only by varying the 
tube--C$_{60}$ distance, but more importantly at fixed distances by i)  changing the 
orientation of the Buckminsterfullerene or ii) rotating one of the 
tubes around its cylinder axis. 
Furthermore, it is explicitely shown that structural relaxation determines qualitatively 
the transmission spectrum of such devices. 
\end{abstract}

Driven by advances in chemical
synthesis,  scanning probe microscopy and break junction techniques,
the seminal idea of using molecular\- scale conductors as active
components of electronic devices~\cite{AR74} has received a new impetus
in recent years \cite{CRRT99,JGA00,Nitz01}. Rectification and 
negative differential resistance could already be demonstrated
on the nanoscale \cite{JGA00}. However, mechanical, electrical or 
electromechanical switching devices and transistors still include components of  
mesoscopic dimensions \cite{JGA00}. \\ 
One of the most 
intensively studied molecules in the field is the fullerene C$_{60}$,
whose electronic transport 
properties were  measured and manipulated by scanning tunneling 
microscopy (STM) techniques \cite{JGSC95}. Subsequently, 
several examples of fullerene-based devices have been investigated both
experimentally~\cite{JG97,PLTM97,ZWWH00,PPLAAM00} and 
theoretically~\cite{JGT98,PS99,TGW01,PPLV01}.  
Mechanisms that have been suggested to control the conductance at a 
single C$_{60}$ molecular junction are either of electromechanical nature (utilizing an 
STM tip to compress the molecule ~\cite{JGSC95,JGT98}) or based on charge transfer  
(controlled by a gate potential in a three-terminal
geometry~\cite{TGW01}). The first mechanism has already been realized experimentally, 
although the theoretical explanation of the observed effect is still  
inconclusive~\cite{JGT98,PS99}, whereas the second approach
runs into the difficulty of  applying a gate terminal at very short length 
scales. Experimentally, however, there is some evidence of
a third mechanism that changes the conductance properties
 of C$_{60}$ by altering the orientation of the  molecule with 
respect to a substrate. This proposed mechanism could explain the observed changes in contrast
of STM pictures of molecular C$_{60}$ layers on gold 
surfaces~\cite{Cetal00,Setal01}.

Other prime targets as
possible building blocks for nanoscale electronics devices are single- and multi-wall 
carbon nanotubes (CNTs) \cite{McEu00}. This owes to their exceptional
electronic and structural properties, that have been extensively studied
over the decade following their discovery.
It has e.\ g.\ been demonstrated that tubes can act as both a wiring 
system \cite{Retal00} and active device 
elements~\cite{Retal00,Metal98,Petal01,WMSS01}. In the latter a
gate voltage was applied to manipulate the conductance.
Recently, however, it has also been shown that if CNTs are placed
on a graphite substrate, the interface resistance can be tuned 
by changing the tube orientation on the graphite plane in- or out-of-registry,
with an atomic force microscope tip \cite{Petal00}. Similar effects have been
studied theoretically for a junction of two CNTs in a $\pi$-electron
tight-binding formulation \cite{BL01}.

In this letter, we propose and investigate an alternative switching device entirely on 
the nanoscale which combines the unique features of both,
C$_{60}$ and CNTs, in a CNT-C$_{60}$-CNT hybrid system
(see Fig.\ \ref{fig1}). The electronic transport properties of this pure-carbon 
electronics setup are studied 
in the Landauer formalism \cite{Datt95} using Green function techniques 
combined with a density functional theory (DFT) 
approach. We show that an unexpected  large variation
of the conductance over three orders of magnitude can be achieved by either
changing the orientation of the
Buckminsterfullerene, or by rotating the nanotubes around the
symmetry axis at a fixed tube-C$_{60}$ distance. In addition, it is  
demonstrated that the conductance of such molecular devices strongly depends 
on the structural relaxation of the junction.

In the Landauer approach, the conductance 
of the system is related to an independent-electron elastic scattering 
problem, with the transmission function $T(E)$ at the Fermi energy 
$E_{\rm F}$ determining the two-terminal, linear-response conductance in the zero
temperature limit 
\begin{eqnarray}
 g=\frac{2 e^2}{h}  T(E_{\rm F}).
\end{eqnarray}
The Green function method is employed to calculate $T(E)$ from
\begin{equation}
T(E)={\rm Tr}[{\bf \Gamma_L G^r \Gamma_R G^a}].
\end{equation}
In a basis representation,
the retarded and advanced molecular Green functions ${\bf G^{r,a}}$
are determined by solving the finite dimensional matrix equation
\begin{eqnarray}
(E{\bf S}-{\bf H} -{\bf \Sigma_{L}}-{\bf \Sigma_{R}}) {\bf G} = {\bf 1},
\end{eqnarray}
which can be derived using a partitioning technique for nonorthogonal bases
\cite{PSRB96}. Here, ${\bf H}$ and ${\bf S}$ are 
the molecular Hamiltonian and the overlap matrix, respectively.  The 
self energies ${\bf \Sigma_{L}}$ (${\bf \Sigma_{R}}$) depend on 
the coupling matrix elements between the scattering region
and the left (right) lead and on the  
lead Green functions. Their imaginary parts
determine the matrices 
${\bf \Gamma_{L,R}}=i[{\bf \Sigma_{L,R}}-{\bf \Sigma_{L,R}^\dagger}]$ \cite{Datt95}. 

\begin{center}{
\begin{figure}[h]
\subfigure{\epsfig{figure=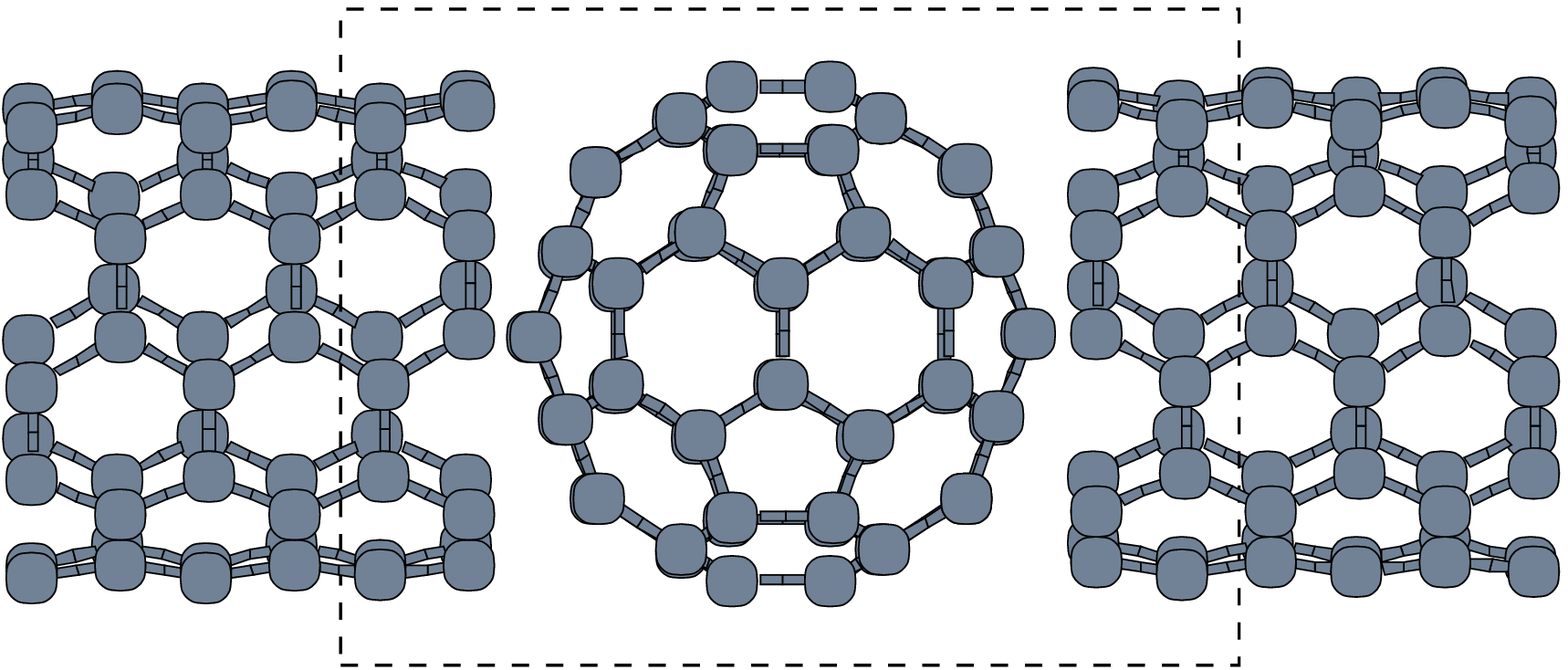 ,width=8cm ,height=3.3cm}}\\
\subfigure{\epsfig{figure=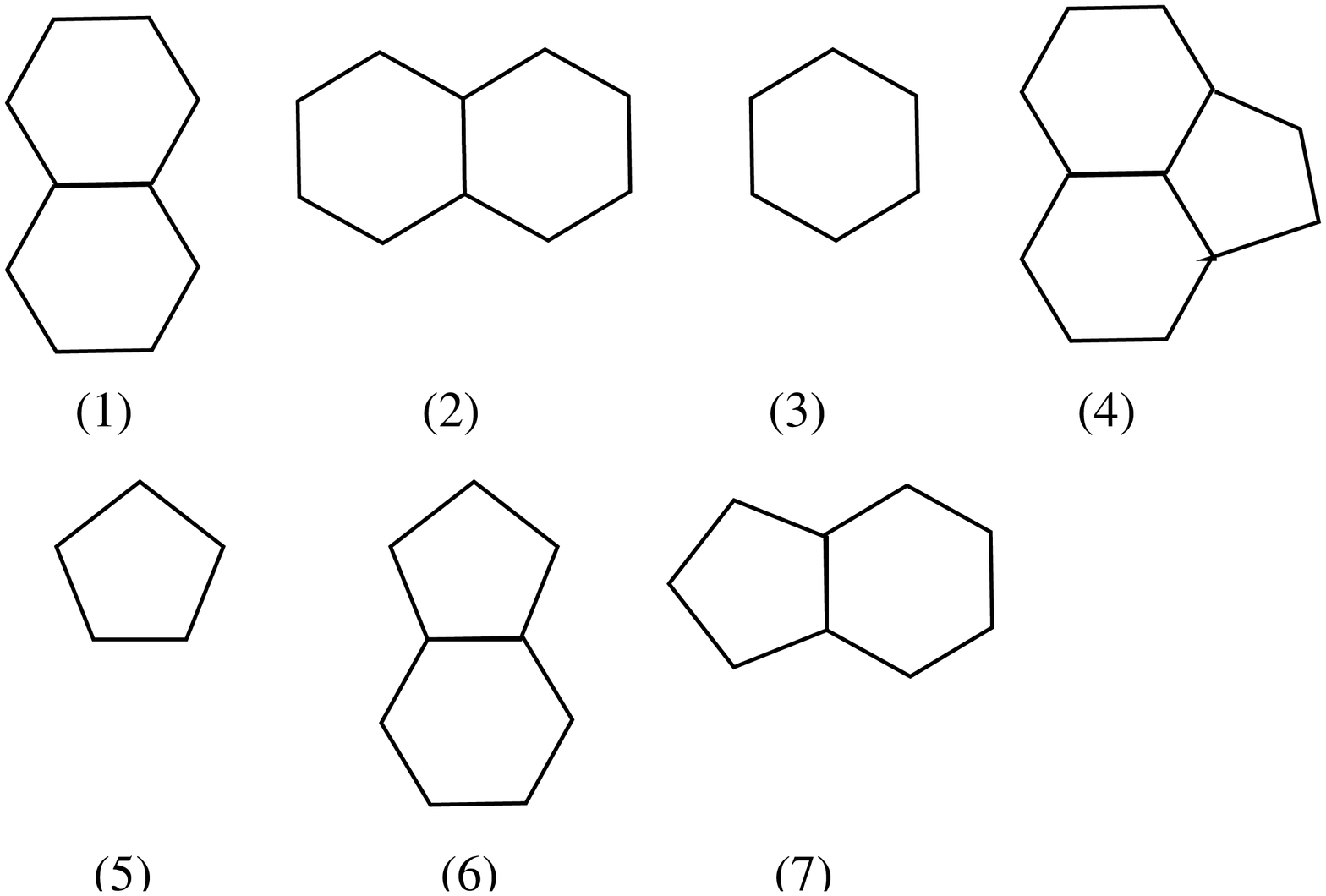 ,height=5cm}}\\
 \caption{\label{fig1}
Geometrical configuration of the molecular junction discussed
in this letter. A fullerene C$_{60}$ molecule bridges two (5,5) CNTs.
The dashed-line frame encloses the scattering region. The lower panel 
represents schematically the different
orientations of C$_{60}$ with respect to the surface cross-sections of the
nanotubes. For example the upper panel geometry corresponds to 
orientation (1).
}
\end{figure}
 }
\end{center}

In the following, we characterize the
electronic structure of the molecule and the leads as well as the
lead-molecule coupling within a parameter-free framework, abstaining
from the use of semiempirical H\"uckel-type approaches. Our method
relies on an approximate DFT based nonorthogonal linear combination of
(valence) atomic orbitals (LCAO) Ansatz, which has been previously applied 
to study dynamical ~\cite{KSS99},   
structural~\cite{PFK95} and  electronic transport \cite{pra01}
properties of a large class of materials. In the present study, at 
each carbon atom an $sp^3$ atomic basis set is located.
The Green functions of the tubes
have been calculated by using a decimation technique \cite{LLR85}. 
Structural optimization is performed by using  conjugate-gradient
techniques and taking into account a cluster consisting 
of the C$_{60}$ cage and six unit cells of the CNT on either side of it. 
To simulate 
the effect of semi-infinite leads we allowed only the fullerene and the first unit cells of the
tubes (nearest to C$_{60}$) to relax, thus defining the scattering region
of Fig.\ \ref{fig1}.

 \begin{figure}[h!]
{\epsfig{figure=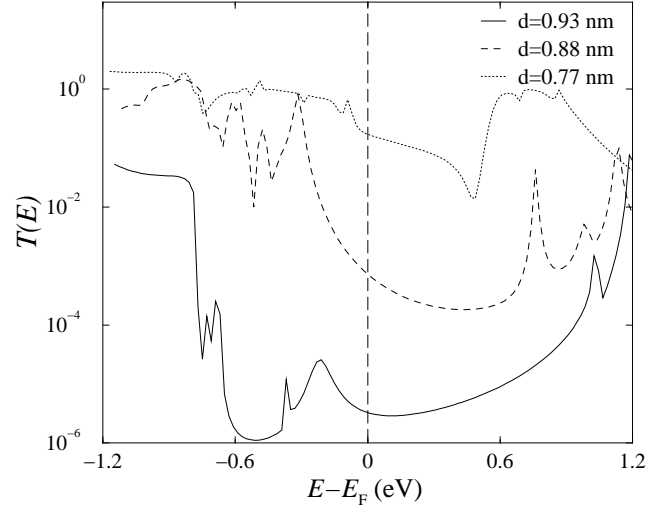 ,height=7cm}}\\
 \caption{\label{fig2}
Transmission spectra for the relaxed configuration  
with orientation (5) (see Fig. 1) for different distances $d$ between 
the nanotubes.}
\end{figure}

We study a single C$_{60}$ molecule 
bridging two single-wall metallic (armchair) (5,5) nanotubes, which are 
taken symmetric with respect to the plane through the center of 
mass of C$_{60}$ and perpendicular to the CNT cylinder axes
(see upper panel of Fig.~\ref{fig1} for a typical unrelaxed structure). 
The tubes act as donor and acceptor electron reservoirs. 
First, we investigate the distance dependence
of the conductance for a given configuration. The central aim of this letter,
however, is to exploit the sensitivity of electron transport to the 
topology of the molecule/electrode interface~\cite{FCR01a} in the
proposed system. To this end
several possible orientations of the C$_{60}$ (depicted by the
polygon(s) facing the tube symmetry axis in the lower panel of Fig.\
\ref{fig1}) have been considered. In addition, the rotation of one of the
tubes around the symmetry axis at a fixed orientation of C$_{60}$
is investigated.
 
The energy dependence of the total transmission at 
three different tube-tube separations $d$ is shown in
Fig.\ \ref{fig2} for the case of the 
(relaxed) configuration (5)  as a function of $E-E_{\rm F}$~\cite{fermi}. 
Large variations in the transmission of up to
5 orders of magnitude are found, similar  to 
those reported in Ref. \cite{PS99} for the case of a
point-like STM-tip approaching a C$_{60}$ molecule.
A reduction of $d$ increases the molecule-lead coupling
and, therefore, leads to a strong broadening and shifting of the
resonances, as can be seen in Fig.\ \ref{fig2}.
For very short separations the cage geometry of C$_{60}$ distorts
and no resonances are  resolved any more.
The molecule becomes highly transparent ($T(E)\sim 1$) over a 
wide energy range around the Fermi level. 
We stress that the presented results for the distance dependence 
are very sensitive 
to the inclusion of structural relaxation and strongly differ from results for 
unrelaxed structures (not shown).
The same effect has been observed for surface induced relaxation
of a CNT-CNT junction in-registry \cite{BL01}.

\begin{figure}[h!]
 {\epsfig{figure=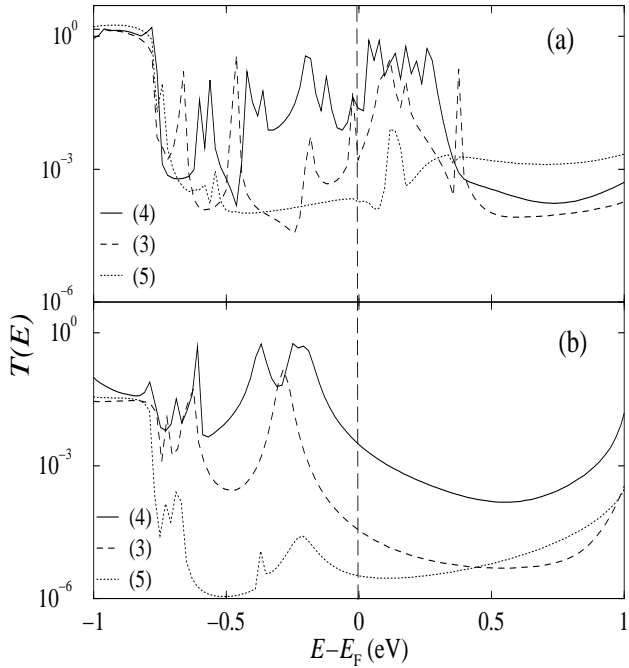 ,width=8.3cm,height=9cm}}\\
 \caption{\label{fig3}
Transmission $T(E)$ as a function of the injected-electron energy.
Results are shown for both unrelaxed (upper panel) and relaxed 
(lower panel) scattering geometries. The tube-tube separation is fixed 
at $d=0.93$ nm. Numbers indicate
different molecular orientations as depicted in the lower panel of
Fig.~\ref{fig1}.
 }
\end{figure}

A key question to ask is what happens
for a fixed distance between the molecule and the tubes
but for different orientations of C$_{60}$ (see lower panel of 
Fig.\ \ref{fig1}) with the inclusion of relaxation.
To study this, we display the transmission as a 
function of the electron energy for both unrelaxed and relaxed 
molecular junction geometries in Fig.\ \ref{fig3}. 
Surprisingly, at fixed distance, just an atomic scale rotation of
the highly symmetric C$_{60}$ molecule induces a large variation of the 
transmission at the Fermi energy by several orders of magnitude.
This is shown in Fig.\ \ref{fig3}(b) for three of the seven different 
orientations with maximum, minimum and one intermediate value of $T(E_{\rm F})$.
As can be seen in Fig.\ \ref{fig3}(a), neglecting relaxation 
decisively influences the transmission 
properties of the molecular junction. This is exemplified as a much 
different and less smooth behaviour of the transmission. In contrast to 
Fig.\ \ref{fig3}(b), no reminiscence of the highest occupied molecular orbital 
(HOMO)-lowest unoccupied
molecular orbital (LUMO) gap is visible in these results ~\cite{super}.
However,
the results for the relaxed structures reveal that,
at the Fermi energy, the pentagon configuration (5) has a transmission
lower by about 3 orders of magnitude than configuration (4). This fact
could be used to build an electronic switching device on the
nanoscale as manipulation of fullerenes by using STM or atomic force
microscope tips is becoming a standard technique in the 
field \cite{ZWWH00,Petal00}.
Furthermore, this effect may also serve to explain the observation of contrast
variations in STM images \cite{Cetal00,Setal01}.

Different from other calculations 
which assume non-carbon electrodes
(aluminum) \cite{TGW01,PPLV01}, the absolute values of the transmission found 
here are
small. In the aluminum case, charge transfer from the electrodes onto
the LUMO pins the Fermi energy 
around this resonance. In our purely carbon-based molecular device, however, 
charge transfer effects are much weaker, as a Mulliken population analysis
shows. Thus, rather than a ballistic-like transport supported by molecular 
states, tunnelling through the HOMO-LUMO gap
is the main mechanism of electronic conductance \cite{MJ97}.
The latter is determined by the superposition of the resonance tails and
for molecules with highly-degenerate HOMO and LUMO manifolds like the
isolated C$_{60}$, an increase of this overlap can be achieved by lifting 
the degeneracy ~\cite{JG97,JGT98}. 
In our calculations, coupling to the leads changes the electronic structure of
the fullerene in a similar fashion. However, our results suggest additional
competing effects.  The wavefunction overlap between components at a molecular
junction which includes nanoelectrodes strongly depends on the
exact configuration \cite{FCR01a} and, hence, the shape and the position
of molecular resonances.
Moreover, the underlying CNT structure
at the electrode-molecule interface is no longer ideal after relaxation and 
this introduces additional resonant states within the HOMO-LUMO gap 
of the isolated  C$_{60}$~\cite{unpub}. 
Note that such states are asymmetrically coupled to the leads,  which explains 
why the value of the conductance at resonances is generally smaller than the quantum unit
($2e^2/h$).

\begin{figure}[h!]
{\epsfig{figure=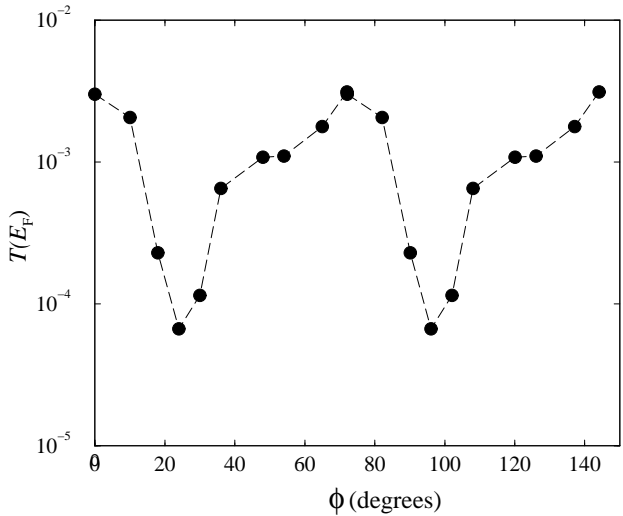 ,height=7cm}}\\
 \caption{\label{fig4}
Transmission at the Fermi energy as a function of the rotation
angle $\phi$  of one of the nanotubes around the direction of 
current transport. The nanotube separation 
is  $d=0.93$ nm and the C$_{60}$
molecule has orientation (4) of Fig.~\ref{fig1}.
   }
 \end{figure}

In addition to the rotation of C$_{60}$ in between the tubes, we have 
also studied
the dependence of the conductance upon rotation of one of the carbon 
nanotube electrodes around the axis of current transport. This destroys the
high symmetry of the CNT subsystem. 
As shown in Fig.~\ref{fig4}, periodic features in the conductance are found and, again,
variations of several orders of magnitude can be observed at the
Fermi energy. 
The rotation angle $\phi=0^\circ$ case 
corresponds to the Fermi level value of the transmission function
for orientation (4) in Fig.\ \ref{fig3}.
From this initial, relatively high conducting situation,
the transmission can be tuned down by two orders of magnitude
by rotating one CNT with respect to the other by $24^\circ$. The period 
of the oscillation
of the conductivity is $72^\circ$, reflecting the symmetry of the tubes.

In conclusion, we have studied a fullerene-based nanobridge of a C$_{60}$
molecule sandwiched between two semi-infinite carbon nanotubes in the
Landauer formalism.  
We have found that in this pure carbon molecular junction the conductance 
is dominated by tunnelling through the
HOMO-LUMO gap. Most importantly, the transmission can be efficiently 
controlled by rotations of the  C$_{60}$ molecule and/or
one of the nanotubes, thus making this carbon hybrid  system
a possible candidate for a nano-electronic switching device. 
Furthermore, we have shown that structural relaxation turns out to have a 
decisive influence on the electronic transport properties
and may not be neglected in this type of calculation.

This research was supported by the ``Deutsche Forschungsgemeinschaft''
through the Forschergruppe ``Nano\-struk\-turier\-te Funk\-tions\-elemente
in makros\-kopi\-schen Sys\-temen''. RG gratefully acknowledges financial 
support by the ``S\"achsische Ministerium f\"ur Wissenschaft und Kunst''.
FG has benefitted from valuable discussions with T. Fritz.


\begin{thebibliography}{10}
\bibitem{fermi}
To determine the
Fermi level for each transport geometry, we have chosen a large
cluster consisting of the C$_{60}$ molecule and several carbon nanotube unit
cells on either side. Then, for increasing
size of the composite, the Hamiltonian matrix was diagonalized and
the electronic states were populated appropriately. The resulting converged
highest occupied molecular orbital (HOMO) was taken as a first 
approximation to $E_{\rm F}$.
\bibitem{super}
Note that the molecule 
is considered to be the `extended' C$_{60}$ previously defined as the scattering 
region ( Fig.\ 1).
\end{thebibliography}
\end{document}